\documentclass[aps,twocolumn,letterpaper]{revtex4}

\usepackage{amsmath}
\usepackage{graphicx}

\newcommand{\dvec}[1]{\ensuremath{\mathbf{#1}}}
\renewcommand{\k}{\dvec{k}}
\newcommand{\q}{\dvec{q}}
\newcommand{\x}{\dvec{x}}

\newcommand{\Tr}{\ensuremath{\mathrm{Tr}}}

\begin{document}

\title{Generation of valley polarized current in bilayer graphene}
\author{D. S. L. Abergel}
\author{Tapash Chakraborty}
\affiliation{Department of Physics and Astronomy, University of
Manitoba, Winnipeg, Canada}

\begin{abstract}
	We propose a device for the generation of valley polarized
	electronic current in bilayer graphene. By analyzing the response of
	this material to intense terahertz frequency light in the presence
	of a transverse electric field we demonstrate that dynamical states
	are induced in the gapped energy region, and if the system
	parameters are properly tuned, these states exist only in one
	valley. The valley polarized states can then be used to filter an
	arbitrary electron current, so generating a valley polarized
	current.
\end{abstract}

\maketitle

\begin{figure}[tb]
	\centering
	\includegraphics[]{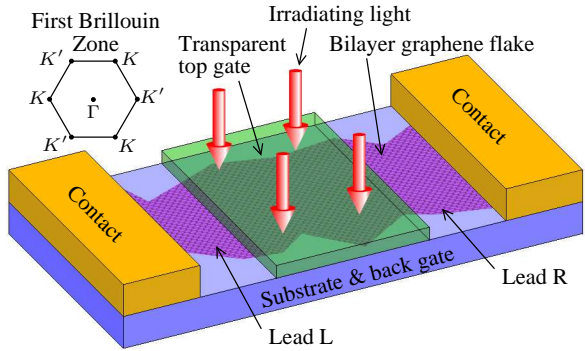}
	\caption{Schematic of the valley filtering device. The area under the
	transparent top gate is irradiated, and the parts of the flake lying
	outside of the gated region function as the graphitic leads. The
	inset shows the first Brillouin zone.
	\label{fig:devsketch}}
\end{figure}

One particularly interesting feature of mono- and bilayer graphene
\cite{Novo-Sci306-NatPhys2, Ohta-Sci313, NovZhang-Nat438} is the valley
degree of freedom. The six corners of the Brillouin zone
(the $K$ points in the inset to Fig. \ref{fig:devsketch}) are separated
from each other in momentum space, and the geometry of the reciprocal
lattice requires that opposite corners are inequivalent so that there
are two species of $K$ point, called `valleys' \cite{McCann-EPJ148}. The
low energy spectrum is localized near the six $K$ points, so that in
this limit, which of the two valleys the electron momentum is located
in becomes a good quantum number. The valley degree of freedom
therefore constitutes a two state system (analogous to the electron
spin) and is often called the `isospin'. This has prompted the
suggestion that the isospin could be manipulated and controlled in a
useful way (so-called `valleytronics'), for example, to make a solid
state qubit \cite{Ryc-NatPhys3_Akh-PRB77}. Of course, in
order to achieve this goal, one must be able to accurately prepare and
manipulate electron states in one valley or another, and to date there
have been several proposals for devices which purport to achieve this
\cite{Ryc-NatPhys3_Akh-PRB77, Tka-PRB79_Cre-PRB77, Xiao-PRL99,
Pereira-JPCM21, Garcia-PRL100, Martin-PRL100}.

Recently, attention has also turned to the optical properties of
monolayer graphene, and its response to linearly and circularly
polarized irradiating light fields has shown interesting features
resulting from the chirality of the electrons and the linear low energy
spectrum \cite{Syz-Oka, Yao-Rod}.

In this Letter, we combine these areas of interest and analyze the
response of the energy spectrum of gapped bilayer graphene
\cite{Ohta-Sci313, McC-Oos-Cas} to external electromagnetic radiation in
the terahertz frequency range. We then propose a device which filters
electrons according to which valley they are in, creating a valley
polarized current.  Specifically, we find that the different sublattice
composition of the wave functions of electrons in opposite valleys
causes them to interact with the irradiating field asymmetrically. When
the radiation and system parameters are properly tuned, dynamical states
existing entirely in one valley are induced. If a current of electrons
in this energy range is passed through the irradiated region, the
absence of available states in one valley means that those electrons are
unable to pass, while electrons in the other valley may. The current
exiting the irradiated region is therefore comprised of electrons in
only one valley, a so-called `valley polarized current'.

This filtering effect is a direct result of the valley asymmetric density
of states in the irradiated region, and is therefore a bulk effect,
independent of the geometry of the sample and its edges.
This gives our device a significant advantage over many prior
proposals as it does not rely on the precise construction of an
edge (as in Refs. \onlinecite{Ryc-NatPhys3_Akh-PRB77,
Tka-PRB79_Cre-PRB77, Xiao-PRL99}), or the exact deposition of a
gate along one crystallographic direction (as in Ref.
\onlinecite{Pereira-JPCM21}), both of which are very challenging tasks.
Reference \onlinecite{Martin-PRL100} also necessitates a complex gating
arrangement to support one-dimensional channels in the graphene.
Even if these devices could be manufactured, the currents they produce
are often only partially polarized, and are localized in one-dimensional
channels, whereas our proposal shows complete valley polarization for
significant current flow in a bulk situation, making the potential for
applications of the current generated by this device much more
plausible.

We model irradiated bilayer graphene using the Hamiltonian $\mathcal{H}
= H_0 + H_U + H(t)$, where $H_0$ is the Hamiltonian of ungated,
unirradiated graphene and $H_U$ represents the inter-layer potential
difference generated by the top gate \cite{McCann-EPJ148}. 
The time dependent term $H(t)$ is the Hamiltonian of the irradiating
field, described by making the Peierls substitution in $H_0+H_U$ with
the vector potential $\dvec{A}=F/|\Omega|\left[ \cos\Omega t, \sin\Omega
t \right]$ (where $\Omega$ is the frequency of the radiation) giving
\begin{equation*}
	H(t) = \frac{\xi v^{}_{F} eF}{|\Omega|} 
	\begin{pmatrix} 0 & 1 \\ 1 & 0 \end{pmatrix}
	\otimes \begin{pmatrix} 0 & e^{-i\Omega t} \\ e^{i\Omega t} & 0
	\end{pmatrix}.
\end{equation*}
The opposite orientation of the circular polarization is employed by
substituting $\Omega \to -\Omega$ in this definition. 
The natural parameter by which to measure the strength of coupling of
the electrons to the field is $x=\frac{v^{}_F eF}{\hbar\Omega^2}$, where
$F$ is the field intensity, $\Omega$ is the frequency of the radiation,
and $v^{}_F$ is the Fermi velocity. If $x>1$ we say we are in the
strongly irradiated regime. We take the dipole approximation and assume
that the graphene is clean enough that we can ignore inter-valley
scattering caused by defects such as lattice imperfections.  
We also neglect electron-electron interactions. 
The time dependent part of the Hamiltonian is periodic with period
$T_0=2\pi/\Omega$, so we can employ Floquet's theorem
\cite{Dittrich-book} to write the electron wave functions $\Psi(t)$ in
the irradiated region as $\Psi_A(t) = e^{-i\varepsilon^{}_At}
\Phi^{}_A(t)$ where the coefficient $\varepsilon^{}_A$ is the energy of
the dynamical state (called the `quasienergy'). The wave function in the
temporal Brillouin zone $\Phi^{}_A(t)$, defined for $-\pi/\Omega < t <
\pi/\Omega$, is periodic in time and can be expanded over its Fourier
components $n$ and the state basis consisting of eigenfunctions of the
static Hamiltonian denoted $\psi^{}_\alpha$. We therefore write
\begin{equation}
	\Psi^{}_A(\x,t) = \sum_{n=-\infty}^{\infty} \sum_{\alpha}
	e^{in\Omega t} \chi_{n\alpha}^A \psi^{}_\alpha(\x).
\end{equation}
We solve the time dependent Schr\"odinger equation for $\mathcal{H}$ by
taking the Fourier transform of the matrix elements of the Hamiltonian
over the states $\psi^{}_\alpha$ and constructing the Floquet matrix
\cite{Dittrich-book}. Diagonalizing this matrix yields the quasienergies
and the wave function coefficients $\chi^A_{n\alpha}$.

\begin{figure}[tb]
	\centering
	\includegraphics[]{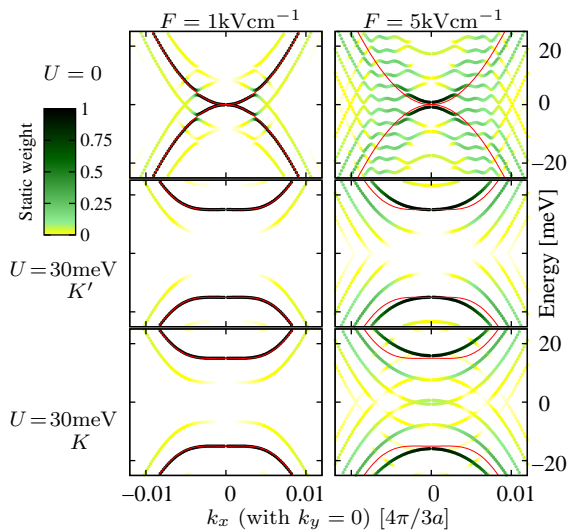}
	\caption{The quasienergy spectrum for $U=0$ (top line) and
	$U=30\mathrm{meV}$ (lower two lines) in both valleys, for weak
	radiation (left) and strong radiation (right). The color of the line
	indicates the weight of the static ($n=0$) component. Thin red lines
	show the unirradiated spectrum.
	\label{fig:qespec}}
\end{figure}

In Figure \ref{fig:qespec} we show the low energy spectrum of irradiated
bilayer graphene with and without a static gap in each of the two valleys. The color
of the line indicates the weight of the static component of the wave
function, which represents the physically observable part of the
dynamical state. In the left-hand column, the coupling parameter
is $x=0.96$ (weakly irradiated) while in the right-hand column $x=4.82$
(strongly irradiated). We superimpose the unirradiated ($F=0$) spectrum
(red lines) for comparison. 
The radiation opens dynamical gaps at $\hbar\Omega/2$ intervals (as was
shown in the monolayer case \cite{Syz-Oka}).  Secondly, when there is a
gate potential applied, dynamical states are present in the gapped
region (see the lower two rows), and the quadratic shape of the low
momentum part of the bands is restored for strong radiation. However,
because $K$ electrons couple more strongly to the radiation than $K'$
electrons (due to the different sublattice composition of the wave
functions), the weights of the static component of the Floquet states
are drastically different in each valley. In the strongly irradiated
regime, the notion of the static gap loses its meaning as there are many
dynamical states with significant static component in that energy range.
It is the dynamical states in the static gap which we utilize in the
proposal for the valley filtering device.  Reversing the polarization of
the light or the orientation of $U$ causes the $K'$ valley to couple
strongly.

\begin{figure}[tb]
	\centering
	\includegraphics[]{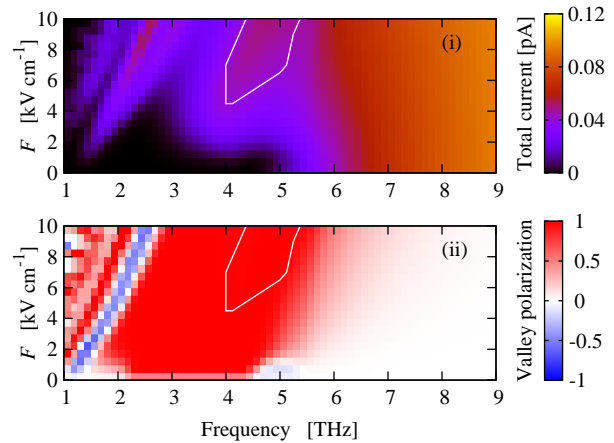}
	\caption{(i) The total current, and (ii) the valley polarization as
	a function of the field parameters. In both plots,
	$U=20\mathrm{meV}$, $\mu=12\mathrm{meV}$, and
	$\eta=0.3\times\hbar\Omega$ with $\Omega=2\mathrm{THz}$. 
	The white contours denote the region of high valley polarization
	\emph{and} significant current flow.
	\label{fig:jomegaf}}
\end{figure}

We now demonstrate the generation of valley polarized current by using
irradiated bilayer graphene as a filter for an arbitrary current. 
We employ a tunnelling approach \cite{Haug-book} where we suppose that
the system consists of three parts, as shown in Fig.
\ref{fig:devsketch}.  They are the two graphitic `leads' described by
Hamiltonians $H_L,H_R=H_0$ with energy spectrum $E_\alpha$ and chemical
potential $\pm\mu/2$, and the central, irradiated region described by
the time dependent Hamiltonian $H_C = \mathcal{H}$ discussed above, with
quasienergy spectrum $\varepsilon_A$ and chemical potential fixed at
zero. The contacts shown in Fig. \ref{fig:devsketch} connect the
graphene flake with external systems, and we do not consider their
influence. The central region is linked to the leads via the coupling
Hamiltonians $H_{CL},H_{CR}$. Denoting the operators for electrons in
the leads by $c_{\k\alpha i}$ for $i\in\{L,R\}$, and the central area by
$d_{\q A}$, we have 
\begin{equation*}
	H_{Ci} = \sum_{\k,\alpha,\q,A} V_{\k\alpha,\q A}
c^\dagger_{\k\alpha i} d^{}_{\q A} + \mathrm{H.C.}
\end{equation*}
We assume that the central region is wide enough to forbid electrons
from tunnelling directly between the two leads.
Since it has been shown \cite{Nilsson-PRB76} that transmission from
bilayer graphene into gapped bilayer graphene is high for a wide range
of the electron's angle of momentum, we assume that for the transfer to
occur, the electron's momentum is conserved and the energy of the
states in the two regions must be sufficiently close. We parameterize
this closeness by writing the function $\Delta(E)$ such that
$\Delta(E)=0$ for $|E|>\eta$ and $\Delta(0)=1$ so that $\eta$
describes the width of the allowed transition.
Then, the coupling parameter is $V_{\k\alpha,\q A} = \mathcal{V}
\delta_{\k,\q} \Delta(E_{\k\alpha} - \varepsilon_{\q A})
\left|\chi_{0\alpha}^A\right|^2$.
The quantity $\mathcal{V}$ has units of energy and parameterizes the
maximal strength of the coupling and we preserve the electron momentum
via the $\delta$ function. 

The valley component of the charge current in the right-hand
lead is $J_\xi = -\langle dN^\xi_R/dt \rangle$, where
$N^\xi_R$ is the number operator for the appropriate electron species.
Using a nonequilibrium Green's function analysis and taking the steady
state limit, we find that the current is
\begin{equation}
	J_\xi = -\frac{2e}{\hbar} \int \frac{d^2\k}{(2\pi)^2}
		\sum_{\alpha^{}_\xi} \Tr \left\{ \bar{\Gamma}_{\alpha^{}_\xi}
		\,\Im \bar{G}^r(E_{\alpha^{}_\xi}) \right\} \mathcal{F}
\end{equation}
where $\Im\bar{G}^r$ is the imaginary part of the full retarded Green's
function in the central region, $\bar{\Gamma}$ contains the coupling
parameters, and $\mathcal{F} =  f_c(E_{\alpha^{}_\xi}) -
f_R(E_{\alpha^{}_\xi})$ depends on the distribution functions in the right
lead and central region.
The central region Green's function is calculated using the
Floquet states derived above, and includes the self energy due to the
two leads.

To characterize the degree of valley polarization of
the current, we define $\mathcal{P} = (J^{}_K - J^{}_{K'})/(J^{}_K +
J^{}_{K'})$ so that $\mathcal{P}=-1(+1)$ corresponds to fully $K'$ ($K$)
polarized current.
In Fig. \ref{fig:jomegaf} we plot the total current and the
polarization as a function of the radiation intensity and frequency for
$U=20\mathrm{meV}$ and the chemical potentials of the leads arranged to
drive current in the energy range corresponding to the static gap
($\mu=12\mathrm{meV}$). The area enclosed by the white contour shows
where $J>0.04\mathrm{pA}$ and $\mathcal{P}>0.98$ simultaneously,
\textit{i.e.} the region where the system parameters are tuned for
significant current \emph{and} very high polarization. 
Reversing the sign of $U$ or the orientation of the polarization of the
radiation leaves Fig.  \ref{fig:jomegaf}(i) unchanged, but inverts Fig.
\ref{fig:jomegaf}(ii) so that the region of high current and
polarization is in the $K'$ valley.
Identification of the valley into which the current is polarized may be
achieved by application of an in-plane electric field \cite{Xiao-PRL99}
which produces a valley-dependent Hall current which will result in a
measurable asymmetry in the electron density across the conducting
channel.

In summary, we have described the measurable characteristics of a
graphene-based valley polarized current generator, where we expect
current of $\sim 0.1\mathrm{pA}$ and valley polarization of $>99\%$. Our
work should provide necessary stimulus in the quest for valleytronics
with graphene.

DSLA thanks H. Schomerus for helpful discussions, and this work
was supported by the Canada Research Chairs programme, and the NSERC
Discovery Grant.


\begin{thebibliography}{}

\bibitem{Novo-Sci306-NatPhys2} K. S. Novoselov, A. K. Geim, S. V.
Morozov, D. Jiang, Y. Zhang, S. V. Dubonos, I. V. Grigorieva, and A. A.
Firsov, Science \textbf{306}, 666 (2004);
K. S. Novoselov, E. McCann, S. V. Morozov, V. I. Fal'ko, M. I.
Katsnelson, U. Zeitler, D. Jiang, F. Schedin, and A. K. Geim, Nat.  Phys.
\textbf{2}, 177 (2006).

\bibitem{Ohta-Sci313} T. Ohta, A. Bostwick, T. Seyller, K. Horn, and E.
Rotenberg, Science \textbf{313}, 951 (2006).

\bibitem{NovZhang-Nat438} K. S. Novoselov, A. K. Geim, S. V. Morozov, D.
Jiang, M. I. Katsnelson, I. V. Grigorieva, S. V. Dubonos, and A. A.
Firsov, Nature (London) \textbf{438}, 197 (2005);
Y. Zhang, Y.-W. Tan, H. Stormer, and P. Kim, 
\textit{ibid} \textbf{438}, 201.

\bibitem{McCann-EPJ148} 
E. McCann, D. S. L. Abergel, and
V. I. Fal'ko, Eur. Phys. J. Special Topics \textbf{143}, 91 (2007).

\bibitem{Ryc-NatPhys3_Akh-PRB77} A. Rycerz, J. Tworzyd{\l}o, and C. W. J.
Beenakker, Nat. Phys. \textbf{3}, 172 (2007);
A. R. Akhmerov, J. H. Bardarson, A. Rycerz, and C. W. J. Beenakker, \prb
\textbf{77}, 205416 (2008).

\bibitem{Tka-PRB79_Cre-PRB77} G. Tkachov, \prb \textbf{79}, 045429
(2009);
A. Cresti, G. Grosso, and G. P. Parravicini, \prb \textbf{77}, 233402
(2008).

\bibitem{Xiao-PRL99}
D. Xiao, W. Yao, Q. Niu, \prl \textbf{99}, 236809 (2007).

\bibitem{Pereira-JPCM21} J. M. Pereira, Jr., F. M. Peeters, R. N. Costa
Filho, and G. A. Farias, J. Phys. Condes. Matter \textbf{21}, 045301
(2009).

\bibitem{Garcia-PRL100} J. L. Garcia-Pomar, A. Cortijo, and M.
Nieto-Vesperinas, \prl \textbf{100}, 236801 (2008).

\bibitem{Martin-PRL100} I. Martin, Ya. M. Blanter, and A. F. Morpurgo
\prl \textbf{100}, 036804 (2008).

\bibitem{Syz-Oka} S. V. Syzranov, M. V. Fistul, and K. B. Efetov
\prb \textbf{78}, 045407 (2008);
T. Oka, and H. Aoki, \prb \textbf{79}, 081406(R) (2009).

\bibitem{Yao-Rod} W. Yao, D. Xiao, and Q. Niu, \prb \textbf{77},
235406 (2008);
F. J. L\'opez-Rodr\'iguez and G. G. Naumis, \prb \textbf{78} 201406(R)
(2008).

\bibitem{McC-Oos-Cas} J. B. Oostinga, H. B. Heersche, X. Liu, A. F.
Morpurgo, and L. M. K. Vandersypen, Nat. Mater \textbf{7}, 151 (2008);
E. McCann, \prb \textbf{74} 161403(R) (2006);
E. V. Castro, K. S. Novoselov, S. V. Morozov, N. M. R. Peres, J. M. B.
Lopes dos Santos, J. Nilsson, F. Guinea, A. K. Geim, and A. H. Castro
Neto, \prl \textbf{99}, 216802 (2007).

\bibitem{Dittrich-book} T. Dittrich, P. H\"anggi, G.-L. Ingold, B.
Kramer, G. Sch\"on, and W. Zwerger \textit{Quantum Transport and
Dissipation}, Wiley-VCH (Weinheim 1998), Chapter 5.

\bibitem{Haug-book} H. Haug, and A.-P. Jauho, \textit{Quantum Kinetics
in Transport and Optics of Semiconductors}, Springer (Berlin, 1998),
Chapter 12.

\bibitem{Nilsson-PRB76} J. Nilsson, A. H. Castro Neto, F. Guinea, and N.
M. R. Peres, \prb \textbf{76}, 165416 (2007).

\end{thebibliography}
\end{document}